# Exploring LLMs Impact on Student-Created User Stories and Acceptance Testing in Software Development

Allan Brockenbrough
Computer Science Department
Salem State University
Salem, Massachusetts, USA
abrockenbrough@salemstate.edu

Henry Feild
Computer Science Department
Endicott College
Beverly, Massachusetts, USA
hfeild@endicott.edu

Dominic Salinas
Computer Science Department
Salem State University
Salem, Massachusetts, USA
d_salinas1@salemstate.edu

## ABSTRACT

In Agile software development methodology, a user story describes a new feature or functionality from an end user's perspective. The user story details may also incorporate acceptance testing criteria, which can be developed through negotiation with users. When creating stories from user feedback, the software engineer may maximize their usefulness by considering story attributes, including scope, independence, negotiability, and testability. This study investigates how LLMs (large language models), with guided instructions, affect undergraduate software engineering students' ability to transform user feedback into user stories. Students, working individually, were asked to analyze user feedback comments, appropriately group related items, and create user stories following the principles of INVEST, a framework for assessing user stories. We found that LLMs help students develop valuable stories with well-defined acceptance criteria. However, students tend to perform better without LLMs when creating user stories with an appropriate scope.

## 1 INTRODUCTION

Large language models based on the transformer architecture and trained on extensive text datasets can complete various tasks, including describing algorithms to solve specific problems, writing coding snippets, and debugging code. The focus on manual coding may decrease with more emphasis on non-coding skills like requirement engineering, design, testing, and integration [1].

Although the code generation skills of LLMs have been an early focus, an LLM tool can also be applied to tasks such as requirement writing, design, code reviews, and testing. We need to understand better when and how an LLM may be helpful in the software development life cycle. Our research investigated the following: (RQ1) Do students working individually, assisted by an LLM through student-directed prompts, create higher quality user stories than when they are unassisted?

## 2 RELATED WORK

In Agile methodology, a user story describes functionality that is valuable to the user and is created through user interviews or feedback [2]. Developers use stories to guide their work by understanding the value provided. We found a limited number of studies researching LLMs and user stories, with even fewer focused on software engineering education. Ronanki et al. [3] studied the quality of software requirements generated by an LLM versus those specified by experts from academia and industry, finding LLMs had higher quality than those generated by experts. In a prior study by the authors [4], we found that students who were given an LLM prompt containing user feedback and using LLM performed better overall than those who did not. For this study, we investigated a question raised by our earlier work: How would the students perform if not given the prompt but had to create it themselves?

## 3 METHODOLOGY

Our study collected 122 user stories from 20 undergraduate computer science students in an upper-level software engineering course at a public university in the United States in spring 2024. The instructor delivered a lecture on the principles of INVEST:

1. *Independent*: Multiple stories should not overlap in content.
2. *Negotiable*: Stories should be flexible and open to discussion with the user without being overly detailed.
3. *Valuable*: Stories should provide clear value to the user and be coherently written to ensure understanding.

4. *Estimable*: Stories should be clear and concise enough for developers to estimate the work required and not overly large.
5. *Small/Scope*: Stories should be small enough to be implemented within several weeks or a Sprint.
6. *Testable/Acceptance Testing*: Stories should include enough information to be testable

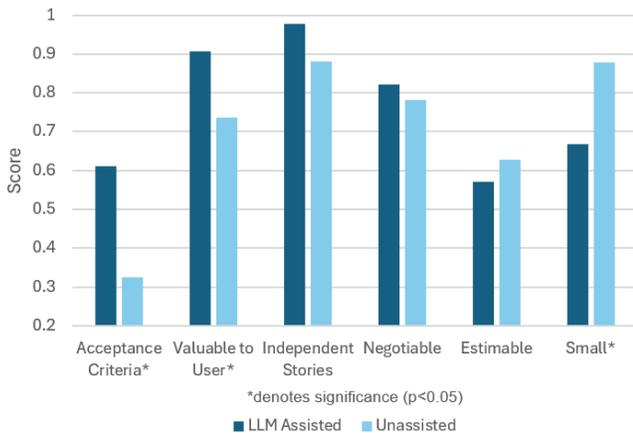

**Figure 1: Average scores for students' user stories attributes.**

We then gave the students an assignment to create user stories for a social media application from user feedback. The assignments were collected, and the students were trained to use an LLM to help build user stories with acceptance testing. The LLM training included (1) creating a separate prompt for each user story with grouped feedback; (2) giving the application context, asking INVEST principles to be followed and acceptance testing generated; (3) analyzing the LLM response, re-prompting if necessary, and manually editing the results. The students were then given a second assignment, with different user feedback from the first assignment, and asked to create user stories with the assistance of ChatGPT 3.5.

### 3.1 Analysis

Two computer science professors independently and blind to LLM usage, scored the depersonalized assignment submissions by assigning a score between 0 and 1 if the INVEST principle had been met. The professors discussed differences in the scoring and agreed upon a single final score for each property.

To address our research questions, we calculated the average score of each INVEST attribute for students using an LLM and students not using an LLM (RQ1). To determine significance, we conducted a paired t-test to compare the mean scores of students who performed tasks with and without the LLM. We used this test to account for the dependency between the two conditions, as each student participated in both tasks.

## 4 RESULTS AND DISCUSSION

The average scores for the attributes of user stories for the subjects (15) that completed both assignments are shown in Figure 1. Among all the attributes, students struggled the most with constructing acceptance criteria without assistance, especially when it came to including sufficient detail for creating test cases. Using an LLM has a statistically significant increase in a student's ability to create acceptance test criteria for a user story. The score of the acceptance criteria increased by 88%. Students using an LLM also had a statistically significant increase in the value of their user stories. To be valuable, a story must be clear and testable. Students using the LLM had an increase of 23% in the scoring of the value attribute.

Using an LLM, however, did not universally improve all attributes. Students who did not use the LLM outperformed those who did in creating stories that were both small enough to be completed within a single sprint and suitable for estimation. When using an LLM students had a statistically significant decrease in their ability to create user stories of the appropriate size. Students using an LLM had a decrease of 24% in the score of the small attribute. There was no statistically significant difference in the scoring for the remaining attributes of independence, negotiable, and estimable.

## 5 CONCLUSION AND FUTURE WORK

This research examines the effects of using LLMs during student generation of user stories from stakeholder feedback in a software engineering course. We described a study in which students were asked to generate user stories for two assignments: one without LLM assistance and one with LLM assistance using a set of instructions. We found that using an LLM significantly improved the ability of the students to create acceptance testing criteria for the story. This suggests that utilizing an LLM for this task could help guide students in creating acceptance tests by offering examples. We also found that the LLM improved the ability of students to create user stories that are valuable. However, students performed better without assistance in creating user stories of the appropriate size. This study highlights the care that must be taken when using an LLM to assist with software engineering development tasks. The use of an LLM does not guarantee better user stories and may lead to inferior results in some attributes. Targeted use of an LLM to assist in crafting acceptance criteria and enhancing the value of the story could be an effective approach. We speculate that providing more guidance on interacting with and modifying the output from an LLM would lead to improvements, particularly the attributes with the most room for improvement, acceptance criteria, and value to the user. Future research should explore this further and expand the sample size to improve the generalizability of the findings.

## REFERENCES

[1] Marian Daun and Jennifer Brings. 2023. How ChatGPT Will Change Software Engineering Education. In Proceedings of the 2023 Conference on Innovation and Technology in Computer Science Education V. 1 (ITiCSE 2023). Association for Computing Machinery, New York, NY, USA, 110–116. https://doi.org/10.1145/3587102.3588815
[2] Mike Cohn. 2004. User Stories Applied: For Agile Software Development. Addison Wesley Longman Publishing Co., Inc., USA.

[3] Krishna Ronanki, Christian Berger and Jennifer Horkoff. 2023. Investigating ChatGPT's Potential to Assist in Requirements Elicitation Processes. In 49th Euromicro Conference on Software Engineering and Advanced Applications (SEAA), 354-361. https://doi.ieeecomputersociety.org/10.1109/SEAA60479.2023.00061

[4] Anonymous (removed for peer review). 2024.